# The CLAVIUS Four Centennial Meeting and XXXI ESOP

Costantino Sigismondi

Galileo Ferraris Institute and ICRA, International Center for Relativistic Astrophysics, Rome Italy

**Abstract:**

The XXXI European Symposium on Occultation Projects will be celebrated in ICRANet center of Pescara from 24 to 27 August 2012 (www.icranet.org/clavius2012). The occasion is the fourth centennial of the Jesuit astronomer Christopher Clavius (Bamberg 1538- Napoli 1612). The hybrid eclipse witnessed by Clavius in Rome (1567) and published on his Commentarius on the Sphere (1581 edition) was the first account of an annular eclipse ever published in a scientific book. To account of this eclipse a larger solar diameter for 1567 has to be considered, and the scientific debate is still open. This is the trait-d'union between Clavius and ESOP annual meeting. The city of Pescara and the region of Abruzzo are presented with an historical, climatic, religious and gastronomical outline.

## Clavius, the Euclid of XVI century

Christopher Clavius (Bamberg 1538- Napoli 1612) was one of the greatest astronomers working at the dawn of telescopic age and contributing to the Copernican revolution. He taught mathematics and astronomy at the Collegio Romano for four decades, earning the title of "The second Euclid" and gave a contribution to the Gregorian reformation of the Calendar (1582) of paramount importance. The hybrid eclipse witnessed by Clavius in Rome (1567) and published on his Commentarius on the Sphere (1581 edition) was the first account of an annular eclipse ever published in a scientific book. According to Ptolemy's parameters such an eclipse was impossible because the angular solar diameter would never be larger than the lunar one. This eclipse was considered by J. Eddy in 1978 in order to demonstrate a larger physical diameter of the Sun before the Maunder minimum (1645-1715). The eclipse project has been carried out by several fellows of the European Section of International Occultation Timing Association (IOTA/ES), and by timing the Baily's beads the solar angular diameter is recovered up to a few hundredths of arcsecond of accuracy. This is the trait-d'union between Clavius, IOTA and solar diameter measurements: a project to monitor the solar diameter with drift-scan methods from ground is named Clavius.

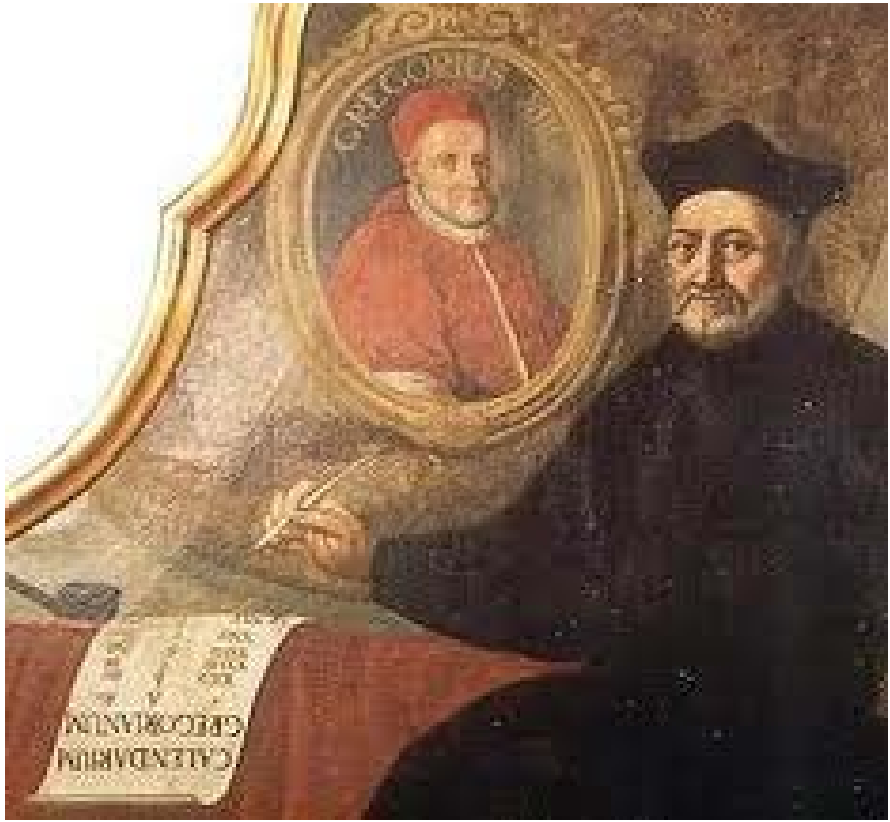

Fig.1 Clavius with the image of pope Gregory XIII, the reformer of the calendar.

A section dedicated to Clavius will have prominent scientists and historians, ready to present this figure of very high level at the dawn of telescopic era in astronomy.

**Occultation astronomy and Relativity**

Occultation astronomy, among all classical astronomy, provides the more accurate measurements of positional and physical parameters of asteroids, TNO and stars, paving the way to all relativistic measurements. That's why the International Center for Relativistic Astrophysics Network coordinating center of Pescara welcomes this meeting. Since more three decades ESOP gathered professional and amateur astronomers to share projects and observations based on the accurate timing of the occultations (asteroidal, lunar and Baily's beads).

Among the works relating Occultation Astronomy and General Relativity I can refer to the references of the following papers: Astrometry and Relativity [1]; Relativistic Implications of Solar Astrometry [2] where the topic of the solar oblateness, a possible issue of eclipse measurements, is strictly related with the Dicke's experiments on quadrupole moment of the Sun in order to explain the anomalous precession of Mercury's perihelion. Also the paper dealing with the occultation of 161 Rhodope over Regulus of 19 october 2005 [3] showed the possibility to observe the relativistic light bending in the gravitational field of the Sun even 48° apart.

The hospitality of ICRANet, an international organism dedicated to General Relativity studies, for the ESOP meeting is therefore grounded also over solid scientific basis.

**Pescara and Abruzzo**

The yearly meetings, started before the fall of the Berlin's wall, were organized with the alternance of Eastern and Western Europe. Pescara, on the Adriatic Sea, is also a natural gate open to the Eastern Countries, and welcomes eagerly the XXXI ESOP in 2012.

Pescara is facing Albania, and the harbor of Split, in the former Jugoslavia. So the influence of Eastern Countries in this city is strong, and the presence of foreign people is normal.

Pescara is a young city, having celebrated its 85 years of foundation on January 2, 2012. But this city is the merge of two former villages, Castellamare and Pescara.

The geographical position allowed to this city to develop rapidly. With about 300000 people leaving in its surroundings Pescara become the most populated urban area of Abruzzo, the region immediately to the East of Lazio, were Rome is. In the years 70s an highway has been realized between Rome and Pescara, the A25 branch of the national highway network, and this allows in 1 hour and half to go from the capital to that pearl of the Adriatic sea.

Abruzzo is called the Green Region of Europe, because agriculture is still the major source of its economy, and was under subventions from EU in the previous decade. It maintained its medieval traditions rather unchanged until the very last years, thanks to the geographical insulation due to the horography. Appennines Mountains separate Abruzzo from Rome, and from the wealthier region of Marche at North, and from Puglia at South, where the shepherds from Abruzzo travelled to spend the winter with the flock. The climate in the mountain region is severe, reaching in some closed valleys the lowest temperatures of the whole peninsula (-32°C). Moreover thanks to the barrier that the mountains offer to the western currents from Mediterranean and from Atlantic, the region is under the influence of Balcan area, with the possibility to experience cold winters (record -13°C on January 4, 1979) even in Pescara.

That is for saying that the people from Abruzzo were used to face hard conditions of work in a life of sacrifices in order to obtain the food from the Earth.

**Religion and history**

Other important aspects of that region, as all regions of Italy, are the religious traditions. Now by the young generations, subjected to the globalization, these traditions are perceived more as touristic attractions, but the ancestral strength of these practices gains power as the youth matures.

In the mountains of Majella and Morrone the hermit Pietro Angelerio settled himself and a community of monk flourished in the fall of XIII century. Later in 1294 while he was in the mountain monastery of Sulmona he was elected pope, and he chose the name of Celestin V. He decided to be crowned in L'Aquila on August 29, 1294 the day of the feast of St. John the Baptist. After his death he was declared saint because of his many miracles.

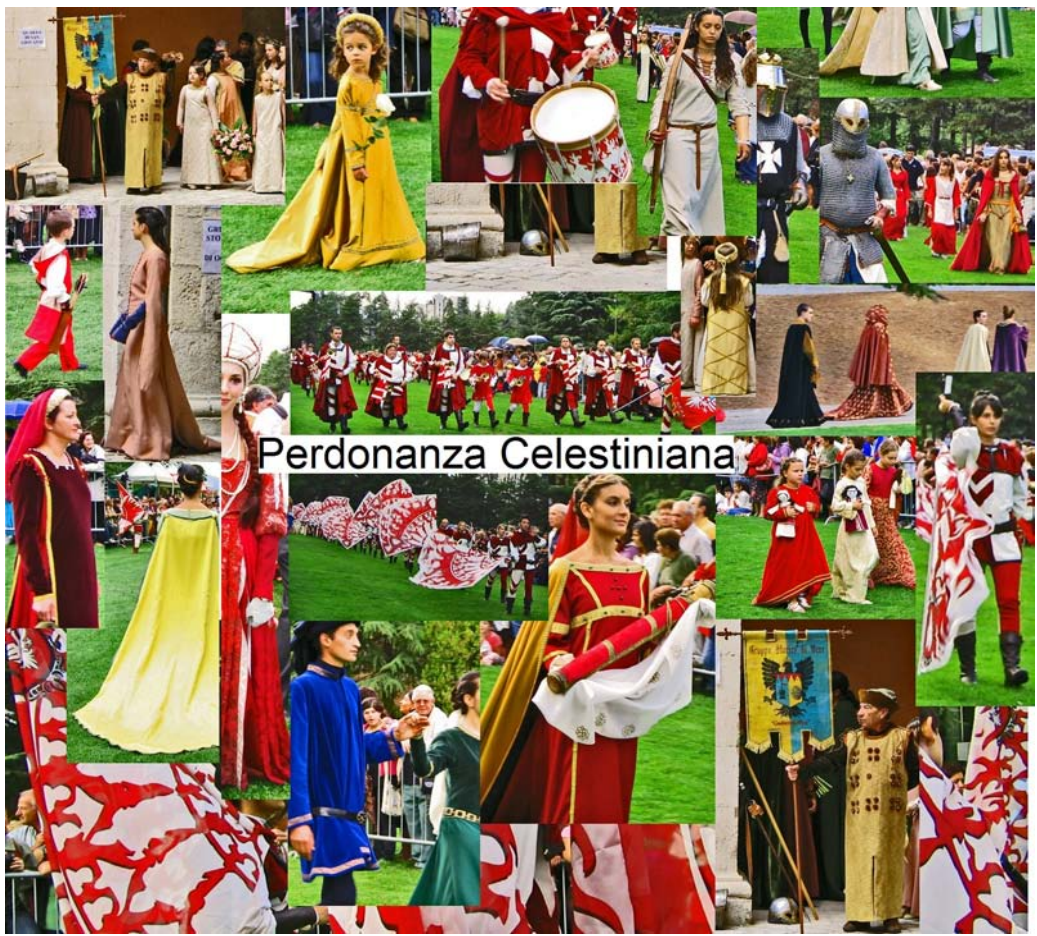

Fig. 2 The Perdonanza Celestiniana in L'Aquila on August 28-29.

This day become the first Jubilee, and the Jubilee of 1300 was inspired by this first event. A great ceremony was celebrated every year in L'Aquila from the evening of 28 to the evening of 29 of August: la Perdonanza. The 718th Perdonanza will occur right after the end of the ESOP meeting, and it is one occasion to visit the city of L'Aquila, the city founded by the "stupor Mundi" Frederic II of Svevia and twice destroyed by an earthquake in 1703 and in 2009. "Immota manet", fixed stays as the motto of L'Aquila says…

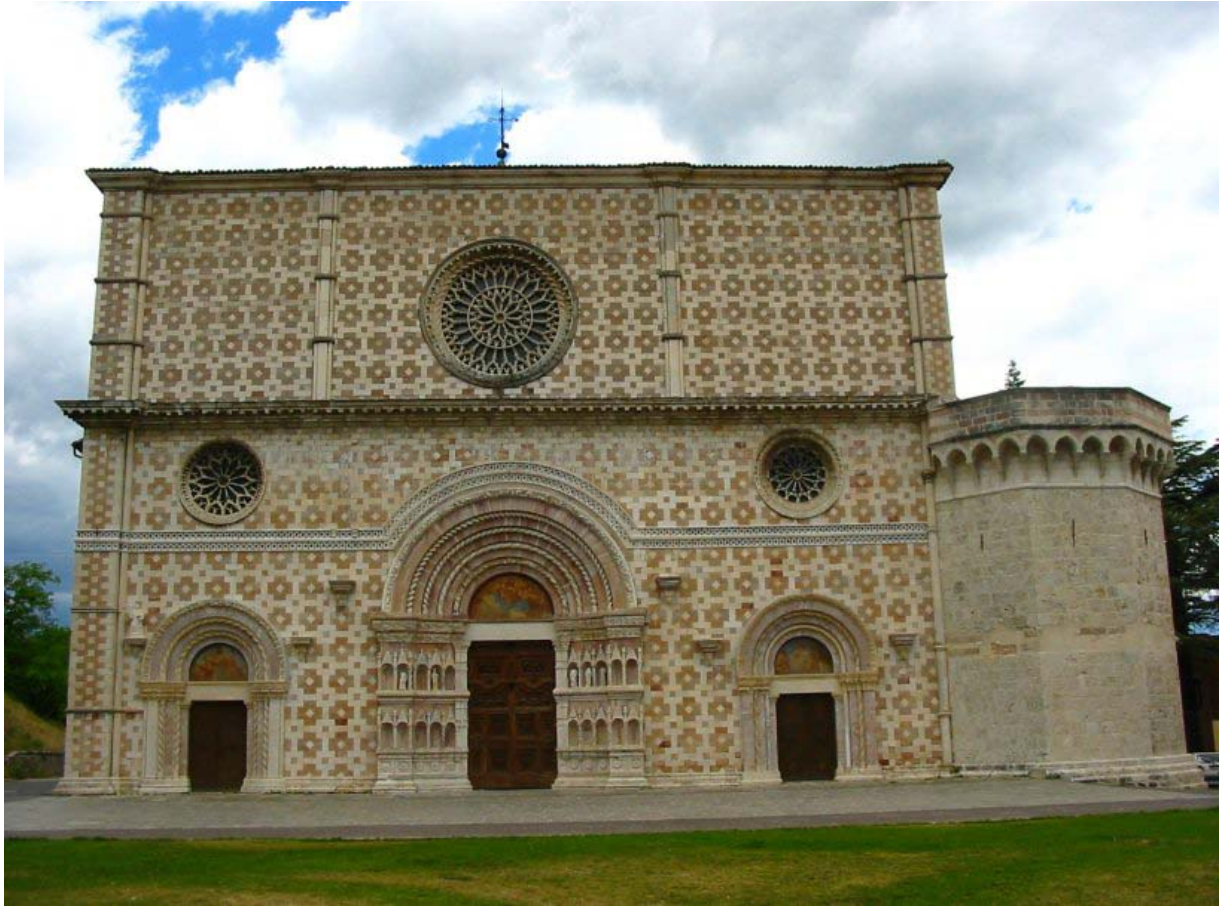

Fig. 3 The church of Santa Maria in Collemaggio, built by saint Pietro Celestino, is one of the marvels of Abruzzo. There is the holy door opened on 28-29 August.

**Summer in Pescara**

The month of August is the warmest of the year. The Foehn wind through the Majella mountain, called Garbino, can raise the temperature of the air up to the record of 45°C of August 30, 2007, with very low humidity. The city is comfortable and the sea breeze makes the afternoon hours rather mild.

Pescara has some free beaches; one of them is 400 meter in front of the location of the congress, in the most central position of the city, very close to all hotels. Other beaches are equipped with all services, and the entrance is upon payment.

Gastronomical tradition in Abruzzo is incredibly rich, and nobody can complain of food in this region, which hosts the most famous school of Chefs de Cuisine in the World: the one of Villa Santa Maria.

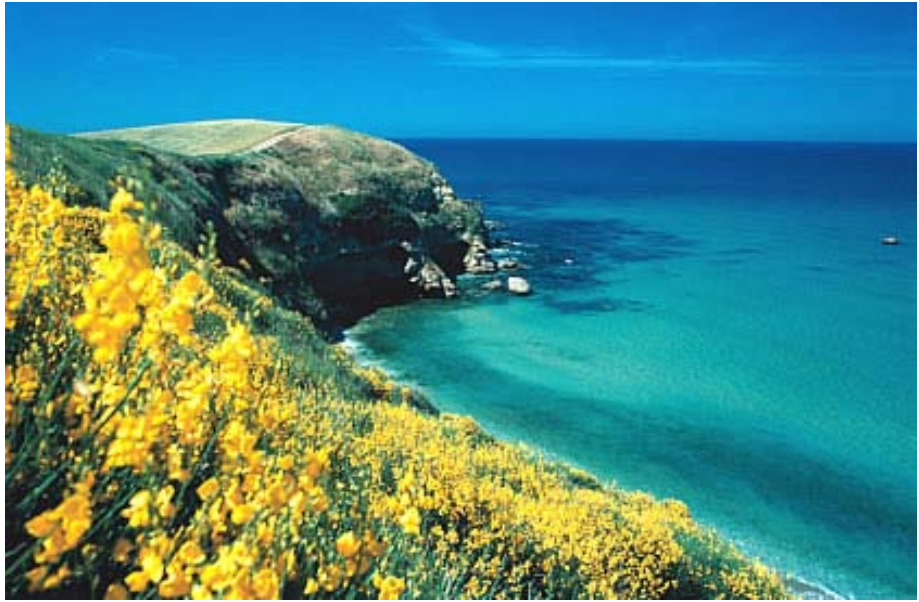

Fig. 4 La costa dei trabocchi, seaside in the surroundings of Pescara.

The invitation to attend the Clavius four centennial and the XXXI ESOP meeting and to know the Green Region of Europe, where also the sea is *green as the pastures of the mountains* (to quote Gabriele d'Annunzio, a famous italian poet born in Pescara) is made. Please go the website www.icranet.org/clavius2012

Fig.5 the poster of the meeting, with the new sea bridge (2009) and the porto-canale, harbor on the river.